\documentclass[conference]{IEEEtran}
\IEEEoverridecommandlockouts
\usepackage{cite}
\usepackage{amsmath,amssymb,amsfonts}
\usepackage{graphicx}
\usepackage{textcomp}
\usepackage{xcolor}

\usepackage{tabularx} 
\usepackage{makecell} 
\usepackage{bm}  
\usepackage{algorithm}  
\usepackage{algpseudocode}  
\usepackage[utf8]{inputenc}  
\usepackage{caption}
\usepackage{subcaption}  
\usepackage{hyperref}  
\hypersetup{
    colorlinks=true,
    urlcolor=blue,
    linkcolor=black,
    filecolor=black,   
    citecolor=black,
    }
\def\BibTeY{{\rm B\kern-.05em{\sc i\kern-.025em b}\kern-.08em
    T\kern-.1667em\lower.7ex\hbox{E}\kern-.125emY}}
\begin{document}

\title{ A Regression Mixture Model to understand the effect of the Covid-19 pandemic on Public Transport Ridership}

\author{\IEEEauthorblockN{Hugues Moreau}
\IEEEauthorblockA{Université Gustave Eiffel\\
Paris, France \\
hugues.moreau.pro@gmail.com } 

\and
\IEEEauthorblockN{Étienne Côme}
\IEEEauthorblockA{Université Gustave Eiffel\\
Paris, France \\
etienne.come@univ-eiffel.fr} 

\and
\IEEEauthorblockN{Allou Samé}
\IEEEauthorblockA{Université Gustave Eiffel\\
Paris, France \\
allou.same@univ-eiffel.fr}

\and
\IEEEauthorblockN{Latifa Oukhellou}
\IEEEauthorblockA{Université Gustave Eiffel\\
Paris, France \\
latifa.oukhellou@univ-eiffel.fr}
}

\maketitle 
\IEEEpeerreviewmaketitle

\begin{abstract}

The Covid-19 pandemic drastically changed urban mobility, both during the height of the pandemic with government lockdowns, but also in the longer term with the adoption of working-from-home policies. To understand its effects on rail public transport ridership, we propose a dedicated Regression Mixture Model able to perform both the clustering of public transport stations and the segmentation of time periods, while ignoring variations due to additional variables such as the official lockdowns or non-working days. 
Each cluster is thus defined by a series of segments in which the effect of the exogenous variables is constant. As each segment within a cluster has its own regression coefficients to model the impact of the covariates, we analyze how these coefficients evolve to understand the changes in the cluster. We present the regression mixture model and the parameter estimation using the EM algorithm, before demonstrating the benefits of the model on both simulated and real data. Thanks to a five-year dataset of the ridership in the Paris public transport system, we analyze the impact of the pandemic, not only in terms of the number of travelers but also on the weekly commute. We further analyze the specific changes that the pandemic caused inside each cluster.

\end{abstract}


\section{Introduction}

Finding similarities between groups of samples, a problem known as clustering, has been an active research domain in the last decades. However, when the samples are time series, one might want to find similarities between different periods. Segmentation, or change-point detection \cite{truong_selective_2020}, is the problem of dividing the temporal axis into intervals in which the data follow a constant distribution. To combine these two approaches, we can use a \textit{sequential} approach, by computing clusters first and finding change points inside each cluster. However, sequential approaches are noticeably unoptimized. Consequently, researchers have developed joint approaches, such as SegClust by Picard \textit{et al.} \cite{picard_segmentationclustering_2007}, or ClustSeg \cite{same_model-based_2011}. 

The introduction of exogenous variables in the mixture model serves two purposes: firstly, it can help the model to remain unaffected by their effects. In the urban mobility domain, it could be interesting to include exogenous variables that represent the three lockdowns in order to enable the model to explain the changes in the data using regression coefficients rather than new segments. Secondly  we want our model to be able to use the influence of these variables to be important to our model: it is worth considering that during typical pandemic behavior, the disparity between weekends and weekdays may be smaller than in the period prior to the year 2020. This is the rationale of the publications which perform a joint regression and clustering \cite{de_veaux_mixtures_1989} or regression and segmentation \cite{liu_segmented_1997, muggeo_estimating_2003}. By explaining the variations in the data with exogenous variables, adding regression to a clustering-segmentation model helps the segmentation to detect only change points for meaningful variations, instead of reacting to variations that can be considered trivial or easily explainable given the context. 

For the analysis of public transport ridership, clustering and regression have already been used successfully either to know which types of travel need improvement \cite{park_machine_2022}, or to determine how investigate one metro line can help alleviate the flow of passengers when another line or station is unavailable \cite{de_nailly_what_2021}. However, these studies focus on signals for which the distribution has the same parameters for the whole duration. The goal in the present paper is to go further and include the effects of the pandemic in the analysis. For instance, to model the difference between weekdays and weekends, we may compute one mean for each type of day \cite{park_machine_2022}, or with an additive model \cite{de_nailly_what_2021}. The latter allows the model to combine the influence of several variables, such as the day of the week and the position in the year. Furthermore, having several segments, each with its regression coefficients, could help us understand how the pandemic changed travel behaviors by looking at which coefficients differ before and after the pandemic crisis. 

This paper aims at better understanding how the pandemic affected the usage of the Paris public transport network. To do so, we used a public transport ridership dataset published by the transport organisation authority called Île-de-France Mobilités (IdFM) which contains records of public transport entries on the rail network of Paris city and its suburbs between 2017 and June 2022. 
The data cover the start of the pandemic, as well as the three lockdowns. Obviously, the official lockdowns had a strong impact on public travel but we are also interested in the impact of the pandemic on travel behaviors. We would expect that, if some stations are relatively unaffected by the pandemic, a clustering-segmentation model could create a cluster for these stations, in which the transition points are located at different dates than the pandemic. 

The present work focuses on the model that allows this interpretation to occur, and its contributions can be summarized as follows: 
\begin{itemize}
    \item we propose a regression mixture model which performs clustering and segmentation, and allows the data to follow a regression model where covariates can be of any type. 
    \item we develop a formulation that models both univariate and multivariate data. 
    \item we evaluate this model against a complete set of baselines, both on synthetic and real data.
\end{itemize}

The rest of this paper is organized as follows: after a literature review, section \ref{sec:model_def} presents both our model and the parameter estimation algorithm. Then, we evaluate our proposed model, on synthetic data (section \ref{subsec:synthetic_data}) and on the public transport ridership dataset in section \ref{subsec:results_real_data}. Finally, section \ref{sec:conclusion} provides some concluding remarks and proposes options for future work. To foster research on this domain, we publish our code at \url{https://github.com/HuguesMoreau/GMM_Clustering_Segmentation}.

\section{Related work}

Our work consists of the intersection of three research domains: clustering, segmentation, and regression. 

As the scope of these domains is very broad, we provide only a concise overview, focusing on their intersections.

Clustering is a field of research aiming to group samples according to their similarity. We will use the formalism of \textit{Gaussian Mixture Models}, it is assumed that the cluster each sample belongs to is a hidden variable, which is estimated using the \textit{Expectation-Maximization} (EM) algorithm \cite{dempster_maximum_1977}.

Like clustering, segmentation is also an active domain of research, that consists in finding intervals of time in which the behavior of the variables is as constant as possible. The temporal ordering allows the formulation of dynamic programming algorithms that guarantee to find a global optimum of the objective function (see Truong \textit{et al.} \cite{truong_selective_2020} for a comprehensive review). Apart from dynamic programming algorithms and their variants, another type of model used to perform the segmentation of time series are Hidden Markov Models \cite{leyli-abadi_mixture_2018}. However, the transitions of Hidden Markov Models cannot always be controlled as precisely as required without additional constraints \cite{nystrup_learning_2020}.

When the segmentation is paired with clustering, using dynamic programming requires re-computing a clustering for each time step (as in \cite{picard_segmentationclustering_2007}) which might make the whole algorithm too long to compute. To prevent this problem, Samé \textit{et al.} \cite{same_model-based_2011} developed a model named \textit{ClustSeg} that performs both clustering and segmentation. To obtain an estimate in a reasonable time, the authors also present the rules to estimate the whole partition using a single EM run. This is the approach we expand by adding the contribution of exogenous variables. Note that the ClustSeg model already projects the data on a basis of functions that can be understood as exogenous variables that are independent of the individuals. This is the reason why Chamroukhi renamed this model Piecewise Regression \cite{chamroukhi_piecewise_2016}, even though the regression is, as in \cite{same_model-based_2011}, only on temporal variables. Our contribution is to allow the incorporation of any type of variable and not only time covariates. 

To account for the exogenous variables, we will use a linear model. De Veaux \cite{de_veaux_mixtures_1989} combined linear regression with clustering to obtain a \textit{mixture of regression}, a model where the mean of each cluster is a linear combination of some other variables. The author also proposed an adaptation of the EM algorithm to estimate the parameters of the model, which is still relatively close to the estimation procedure of classical mixture models.   
Another family of models that falls in the intersection between clustering and regression consists in using use a model named Latent class clustering-based random parameter ordered logit model (LCROL, \cite{chang_injury_2021}), where the probability for any individual to belong to a cluster follows a logistic regression, depending on the provided covariates. This idea is similar to the segmentation used by Samé \textit{et al.} \cite{same_model-based_2011} and Chamrouki \cite{chamroukhi_piecewise_2016}, except that the only variable available in the logistic regression is time (for further details, see section \ref{subsec:model_def}).  

The last field of interest is the intersection of segmentation and regression, named \textit{segmented regression}. However, this name often designates studies where the breakpoint is fixed before the start of the analysis. This kind of segmented regression has been used in medicine \cite{bernal_interrupted_2017}, or for the evaluation of public policies \cite{nistal-nuno_segmented_2017}. 
Most of the time, only two segments are used to know whether the difference between the two is significant.

In our case, we will exhibit the results our model obtains on a public transport dataset. But first, we need to present the formulation of our model, along with the formulation of the EM algorithm to estimate the parameters of the mixture model.

\section{Model definition and parameter estimation}  \label{sec:model_def}

\subsection{Definitions and notations}
The data to be analyzed in this article consist of a set of time series observed over the same time grid indexed by $t$. 

\begin{figure}
    \centering
    \includegraphics[width=9cm]{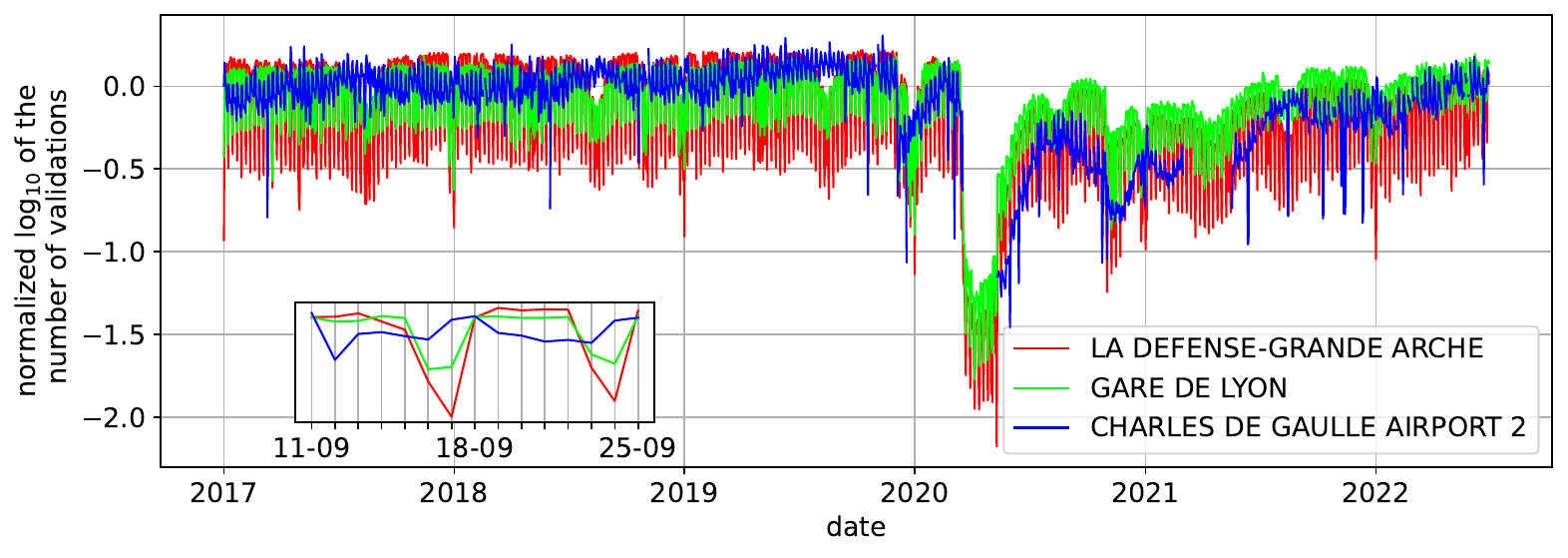}
    \caption{Some examples of series in the dataset we process. The zoom-in represents two weeks starting September 11, 2019 (on a Monday).}
    \label{fig:IdFM_examples}
\end{figure}

The data we want to model are a set of \textit{individuals}, which we want to group into $K$ \textit{clusters}. The dataset covers a time series observed over discrete timesteps $t$ (in our case, days). For each individual $i$, and each day $t$, the data is a $D$-dimensional vector noted $\bm{y}_{i,t}$. Note that the model does not assume that all couples of individuals and days are present: some may be missing from the dataset (and will be in the case of the Public Transport dataset, see fig. \ref{fig:IdFM_examples}). In all cases, we want to obtain, for each cluster $k$, a series of $S$ \textit{segments}, that is, an interval of timestamps with similar behavior. However, there are several causes for change: in the data which we want to exclude from our analysis. This is why we include \textit{exogenous variables} that will help to explain the changes in the observed variables without resorting to new segments. We note $x_{i,t,l}$ the value of the $l^{th}$ variable for individual $i$ and timestep $t$. We assume that all variables are continuous: if one variable has more than two categories, we break it down into several dummy (zero-one) variables. Note that in practice, the variables we will use are mostly constant along all timesteps, or for all individuals.

\subsection{Model definition} \label{subsec:model_def}

Let $z_i$ be the cluster assignment of individual $i$, and $w_{i,t}$ the segment of individual $i$ at time $t$. We assume that the cluster assignments $z_i$ are i.i.d., following a multinomial distribution whose weights are parameters of our model:

\begin{equation}
    P(z_i=k) = \pi_k  \quad \textrm{s.t.} \quad \sum_{k=1}^K \pi_k = 1.
\end{equation}

Given the cluster $k$ of individual $i$, we assume that the segment assignment $w_{i,t}$ follows a multinomial law defined by the probabilities:

\begin{equation} \label{eq:segment_prior} 
    P(w_{i,t}=s|z_i=k) = \kappa^{k,s}_{t} = \frac{\exp(t.u_{k,s} + v_{k,s})}{\sum_{s'}{\exp(t.u_{k,s'} + v_{k,s'})}},
\end{equation}
where $u_{k,s}$ and $v_{k,s}$ are model parameters.
In other words, the only reason why we see definitive transitions is the fact that the arguments of the softmax functions are monotonous. This way of modelling segments using independent day assignments might seem unusual. We use the formulation in equation \ref{eq:segment_prior} because the independence of the segment assignments allows us to make use of the EM algorithm (see section \ref{subsec:EM}). 
The partition of the time axis resulting from the application of the \textit{Maximum a Posteriori} (MAP) rule may not be a series of contiguous segments. We thus applied the following constraints to make sure that the distributions of the temporal class $\kappa_{t}^{k,s}$ vary faster than a given threshold $\lambda$:

\begin{equation} \label{eq:segment_prior_constraint}
    \forall k, \forall s\in \{1 \ldots S-1 \}, \quad  u_{k,s+1} - u_{k,s} > \lambda
\end{equation}

Note that $\lambda$ is not a parameter of our model, but a value we choose before any parameter estimation. Numerically, we set it in such a way that the temporal classes distributions $\kappa_{t}^{k,s}$ go from $0.01$ to $0.99$ in less than three months\footnote{To do so, we solve $\exp({\lambda.t + c}) = 0.01$ and $\exp({\lambda(t + \Delta t) + c}) = 0.99$, yielding $\lambda = (\log(0.99/0.01))/{\Delta t}$. As the time $t$ varies between zero and one in all our formulas, the corresponding ${\Delta t}$ for three months is equal to 90 days divided by the number of days in the interval $T$.}.
We found these constraints paramount: without them, the distribution of temporal classes $\kappa_{t}^{k, s} $ would remain nearly constant (equal to approximately $1/S$, where $S$ is the number of segments), and the posteriors would not have the contiguity expected from segments.

\subsubsection*{Observation model}
Similarly to ClustSeg \cite{same_model-based_2011}, our observation model is a Gaussian linear regression model:

\begin{equation}
    \bm{y}_{i,t}|(z_i=k, w_{i,t}=s) \sim  \mathcal{N}(\bm{\mu}_{i, t}^{k,s}, \bm{\Sigma}_{k, s} )
\end{equation}

However, contrary to ClustSeg, in which the statistical mean $\mu $ depended only on the cluster, segment, and time covariates, our formulation introduces exogenous variables that vary for both individuals $i$ and days $t$. As in mixtures of regressions \cite{de_veaux_mixtures_1989}, the mean is a sum of the linear contributions of all exogenous variables: 

\begin{equation} \label{eq:contribution_exogenous}
    \bm{\mu}_{i, t}^{k,s} =  \bm{m}_{k,s} + \sum_{l=1}^L{x_{i,t,l}.\bm{\alpha}_{l,k,s}},
\end{equation} where $ \bm{m}_{k,s}$ is the intercept of the linear regression model, and the $\bm{\alpha}_{l,k,s}$ are the contributions of the $L$ temporal variables. 

For the covariance matrix of cluster $k$, segment $s$, $\bm{\Sigma}_{k, s}$, we choose to have a diagonal covariance (meaning that all dimensions are independent) but other constraints are possible (full covariance, spherical, \textit{etc.}), similarly to the fact that one can choose the type of covariance for classical Gaussian Mixture Models.

To summarize this section, the free parameters of the model, denoted as $\bm{\theta}$, are the mixture proportions $(\pi_k)_k$, the segmentation parameters $(u_{k,s}, v_{k,s})_{k,s}$, the mean $\bm{m}_{k,s}$ of cluster $k$ and segment $s$, the contributions of the exogenous variables $(\bm{\alpha}_{l,k,s})_{l,k,s}$, and the covariance matrices $(\bm{\Sigma}_{k,s})_{k,s}$. 
The next section explains how to estimate them via the EM algorithm \cite{dempster_maximum_1977}.

\subsection{Parameter estimation} \label{subsec:EM}

The parameter estimation is done by maximizing the log-likelihood:

\begin{align} \label{eq:log_likelihood} 
\mathcal{L}(\bm{\theta}) &=  \log P( \bm{Y}|\bm{X},\bm{\theta}) &\\ 
 &= \sum_{i} \log \sum_{k}  \prod_{t}  \pi_k \sum_{s} \kappa_t^{k,s}  \mathcal{N}(\bm{y}_{i,t};\bm{\mu}_{i, t}^{k,s}, \bm{\Sigma}_{k, s} )  \nonumber    &
\end{align}

The formulation of our model allows us to use the Expectation-Maximization algorithm \cite{dempster_maximum_1977}, a powerful algorithm aiming to estimate the parameters of latent (hidden) variable models. In our case, the latent variable is the union of the cluster and segment, which we estimate using both the observed variable ($\bm{y}_{i,t}$) and the exogenous variables ($\bm{x}_{i,t}$). The EM algorithm consists in alternating two steps.

\subsubsection*{E-step}

In this step, we compute the posterior distribution of the latent variables, using the values of the parameters at the current iteration. As mentioned in \cite{same_model-based_2011}, this is done in two steps. First, we need to compute the posterior probability of a given individual $i$ to belong to cluster $k$, which we note $\rho_i^k$ :

\begin{align} \label{eq:posterior_cluster_proba}
\rho_{i}^{k} &=  P(z_i=k|\bm{y}_{i},\bm{x}_{i}; \bm{\theta})\\
&= \frac{   \pi_k \prod_{t} \sum_{s} \kappa_{t}^{k,s} \mathcal{N}(\bm{y}_{i,t};\bm{\mu}_{i, t}^{k,s}, \bm{\Sigma}_{k, s})  } 
{ \sum_{k'} \pi_{k'} \prod_{t} \sum_{s} \kappa_{t}^{k',s} \mathcal{N}(\bm{y}_{i,t};\bm{\mu}_{i, t}^{k',s}, \bm{\Sigma}_{k', s})   }, 
\end{align}
where we note $\bm{y}_{i} = (\bm{y}_{i,t})_{t \in \{1,\ldots,T\}}$. 

Then, we can obtain the posterior probability for any day $t$ to belong to segment $s$:

\begin{align}
r_{i,t}^{k,s}  &= P(w_{i,t}=s,z_i=k|\bm{y}_{i},\bm{x}_{i}; \bm{\theta}) \\
&= \rho_{i}^{k} \frac{\kappa_{t}^{k,s} \mathcal{N}(\bm{y}_{i,t}; \bm{\mu}_{i,t}^{k, s}, \bm{\Sigma}_{k, s})}
{\sum_{s'}\kappa_{t}^{k,s'} \mathcal{N}(\bm{y}_{i,t};\bm{\mu}_{i, t}^{k,s'},  \bm{\Sigma}_{k, s'}) } 
\end{align}

\begin{equation}
    r_{i,t}^{k,s} = P(z_i=k, w_{i,t}=s|\bm{y}_{i}, \bm{x}_{i}; \bm{\theta}),
\end{equation}

Once we have computed the responsibilities, we update the parameters during the \textit{Maximization} step. 

\subsubsection*{M-step}

In the maximization step, we use the responsibilities $r_{i,t}^{k,s}$ computed earlier, and obtain the parameters that maximize the expectation of the log-likelihood. Conveniently, we can write it as a sum of terms which can be optimized separately:

\begin{align} \label{eq:optim_M_step}
 \mathcal{Q}(\bm{Y}, \bm{x}; \bm{\theta}) &=  \sum_{i,t,k,s} r_{i,t}^{k,s} \log \pi_k \nonumber &\\  
   &+ \sum_{i,t,k,s} r_{i,t}^{k,s} \log \frac{\exp(t.u_{k,s} + v_{k,s})}{\sum_{s'}{\exp(t.u_{k,s'} + v_{k,s'})}}    &\\ 
   &+ \sum_{i,t,k,s} r_{i,t}^{k,s}  \log \mathcal{N}(\bm{y}_{i,t};\bm{\mu}_{i, t}^{k,s}, \bm{\Sigma}_{k, s} ) \nonumber
\end{align}

Thus, each set of parameter values can be computed independently from the others, the only exception being the variance which requires the segment mean and variable contributions of the current step. In this section, we display only the results of the maximization step. 

\paragraph{cluster proportions} As with many mixture models, the mixture proportions are the proportion of the responsibilities of the samples. For each cluster $k$: 

\begin{equation}
    \pi_k = (\sum_{i,t,s} r_{i,t}^{k,s}  )/(\sum_{i,t,k',s} r_{i,t}^{k',s} )
\end{equation}

\paragraph{segment borders} The segment parameters $u, v$ are found by computing the maximum of the following logistic regression problem:

\begin{equation}
    \sum_{i,t,k,s} r_{i,t}^{k,s} \log \frac{\exp(t.u_{k,s} + v_{k,s})}{\sum_{s'}{\exp(t.u_{k,s'} + v_{k,s'})}}
\end{equation}

This problem is convex (even with the constraints mentioned in equation \ref{eq:segment_prior_constraint}), and we solve it using scipy's \texttt{optimize} package implementation of the quasi-Newton method from \cite{conn_trust_2000}. We ensure identifiability by setting $\sum_s u_{k,s} = \sum_s v_{k,s} = 0$ for all $k$.

\paragraph{mean, regression coefficients, and covariance} We can find the optimal values for the last term of equation \ref{eq:optim_M_step} separately for each segment. For all $k,s$, the values of $\bm{m}_{k,s}$, $\bm{\alpha}_{k,s}$, and $\bm{\Sigma}_{k,s}$ minimize:

\begin{equation} \label{eq:exogenous_var_contrib_computation}
    \sum_{i,t} r_{i,t}^{k,s}  \left((\bm{y}_{i,t} - \bm{\mu}_{i, t}^{k,s}) \bm{\Sigma}_{k,s}^{-1} (\bm{y}_{i,t} - \bm{\mu}_{i, t}^{k,s})' + \log{\det  \bm{\Sigma}_{k,s} } \right),
\end{equation}
where $\bm{x}'$ denotes the transpose of vector $\bm{x}$ and $\bm{\mu}_{i, t}^{k,s} = \bm{m}_{k,s} + \sum_{l}{x_{i,t,l}.\bm{\alpha}_{l,k,s}} $. This is a common linear regression problem where the $r_{i,t}^{k,s}$ are weights \cite{rasmussen_gaussian_2004}.

As mentioned earlier, we use these formulas during each of the maximization steps to estimate new values of the parameters. The complete algorithm can be summarized as follows:

\begin{algorithm}
\caption{the EM algorithm}
\begin{algorithmic}

\Require observed data $(\bm{y}_{i,t})_{i,t}$, exogenous variables $(\bm{x}_{i,t})_{i,t}$, a number of clusters $K$ and segments $S$, and an initial parameter $\bm{\theta} $

\While{the log-likelihood has not converged}
    \State \textit{E-step}: Compute the expectation of the latent variables $r_{i,t}^{k,s}$ using the parameters of the model $\bm{\theta}$
    \State \textit{M-step}: Find the parameters of the model $\bm{\theta}$ that best explain the data
\EndWhile
\end{algorithmic}
\end{algorithm}

We keep on alternating the expectation step and maximization steps until the log-likelihood stops increasing by more than $10^{-4}$ in ten iterations. Only then is parameter estimation over and we can start evaluating our model.

\subsubsection*{Initialization}

To initialize our algorithm, we start by setting the mixture proportions equal to $1/K$, where $K$ is the number of clusters. The segment parameters $u,v$ are set in such a way as to respect the constraints we set (see equation \ref{eq:segment_prior_constraint}), and the borders between segments delimit segments of equal duration. The segment means $\bm{m}_{k,s}$ are estimated by assigning each individual a random cluster and each day its corresponding segment, and computing the mean of values inside each segment. Finally, we end the initialization by setting the contribution of exogenous variables to zero.

\section{Experiments}

This section is devoted to presenting the experimental results that highlight the efficacy of the proposed model. We will compare the performance of our model against a set of baselines using artificial data. Subsequently, we will assess its performance on the public transport ridership dataset through cross-validation. Lastly, we will analyze the results obtained by the model to gain insights into the stations that experienced the most significant impacts due to behavioral changes resulting from the pandemic.

\subsection{Protocol}

The model presented above allows us to cover clustering, segmentation, and regression at once. To justify its use despite the additional complexity, we compare it to several alternatives. Most of these methods consist in using ClustSeg \cite{same_model-based_2011} without exogenous variables, which means that the basis of constant functions is used to represent the time series. Conversely, as the proposed model includes covariates in the parameter estimation, the basis of functions we choose to represent the time series can be found in the variables we have (in our case, splines of degree 2). 
To demonstrate the interest of simultaneously performing the segmentation and the clustering, we include in our comparison two methods that compute clusters before segments:
\begin{itemize}
    \item \texttt{Reg $\rightarrow$ Clust $\rightarrow$ Seg}: This method operates in three sequential steps. Firstly, we estimate regression coefficients ($\bm{\alpha}$, see eq. \ref{eq:contribution_exogenous}) common to the whole dataset. Secondly, we perform a clustering of the residual time series derived from the first step. Thirdly, we perform a segmentation  of the clusters obtained in the second step. 
    Note that for synthetic data, we generate the ground truth regression coefficients with uniform distributions. Hence, computing regression coefficients on the whole dataset does not make much sense. In other words, we expect this method to be completely irrelevant for synthetic data. However, it might prove somewhat useful for real data, when the contributions of the exogenous variables might be similar across clusters and segments. We include it nonetheless for both types of data. 
    \item \texttt{(Clust+Reg) $\rightarrow$ (Seg+Reg)}: This baseline involves fitting a mixture of regression model, before performing a segmented regression inside each cluster. It is important to note that we estimate new variable contributions during both the clustering and segmentation processes. 
\end{itemize}

We also include three baselines where the clusters are estimated at the same time as the segments, meaning that the model can use the change points in the samples to estimate how close all individuals are to each cluster: 
\begin{itemize}    
    \item \texttt{Clust+Seg}: We simply reuse the method from \cite{same_model-based_2011}, without any covariates. As ClustSeg requires a basis of (temporal) functions, using it without any covariates is equivalent to determining piecewise constant cluster prototypes. The rationale for including this method is mainly to demonstrate the relevance of including the exogenous variables with real data.    
    \item \texttt{Reg $\rightarrow$ (Clust+Seg)}: This model is similar to the one we propose, except that we assume that the linear regression parameters are constant across all clusters and segments. Note that to estimate the parameters for this model, we perform a linear regression on the whole dataset (before any affectation onto clusters and segments), before using ClustSeg (without any exogenous variables) on the residuals. This estimation procedure is equivalent to estimating the parameters of the model using the classical EM algorithm based on Gaussian mixture models. Similarly to \texttt{Reg $\rightarrow$ Clust $\rightarrow$ Seg}, we expect the common regression coefficients to be ill-adapted to synthetic data. 
    \item \texttt{(Clust+Seg) $\rightarrow$ Reg}: This method involves first using ClustSeg \cite{same_model-based_2011} without any covariables, then performing a regression on each segment. It has the same set of parameters as the one we propose, the only difference between the two being that the model we present has access to changing regression coefficients during the clustering and segmentation. Given that both this method and the first one (\texttt{Clust+Seg}) compute the clusters and segments without any variables, they assign the same partition to the data, which is why we do not use it on synthetic data, where the evaluation criterion is the relevance of the partition. 
\end{itemize}

Using different evaluation criteria, we compare the partitioning obtained with our model to the ones resulting from these methods. For synthetic data (simulations), the creation of the dataset allows us to have access to the latent variables that generated the data as ground truth (cluster and segment of each individual and day). We consider that two samples belong to the same partition if and only if their cluster \textit{and} segment both match, and use the Adjusted Rand Score to compare to the evaluated models' estimations. 

For real-life data, there is no such reference to use. Hence, we look at the log-likelihood on an unseen dataset: we separate the dataset into two subsets for training and validation. We use the training dataset to estimate the parameters of the model. Then, we calculate the log-likelihood of the model by applying it to the validation dataset. The higher the log-likelihood, the better the model explains the validation data, and the better we consider our model.

\subsection{Validation of the model on synthetic data} \label{subsec:synthetic_data}

\paragraph*{Parameter generation}
Before generating data, we need to fix the model parameters. We select arbitrarily $K=4$ clusters and $S = 4$ segments. The mixture proportions $\pi_k$ are drawn from a Dirichlet distribution with parameter $2$, to make sure all clusters appear for at least one sample. Similarly, for each cluster, we draw segment proportions with this same distribution, where all slopes $u$ are set so that a probability changes from $10 \%$ to $90 \%$ in one-tenth of the time interval. We sample the cluster and segment means $m_{k,s}$ with $\mathcal{N}(0,1)$. To imitate the exogenous variables from the real data, we use three exogenous variables: the first is constant across timesteps ($x_{i,t,1} = y_{i,1,1}$), the second is constant across individuals ($x_{i,t,2} = x_{1,t,2}$), and the third varies for both individuals and timesteps. All are drawn from $\mathcal{N}(0,1)$. Then, we sample the contributions corresponding to these exogenous variables ($\alpha$ coefficients) from $\mathcal{N}(0,\bm{\Sigma}_{\alpha})$, where the value of $\bm{\Sigma}_{\alpha}$ changes between experiments. Finally, we generate the observed variables using the variance $\Sigma_{k,s} = 1$. The choice of a relatively high unexplained variance $\Sigma_{k,s}$ is voluntary, to place the model in a difficult scenario.

During the first set of experiments, we set $\bm{\Sigma}_{\alpha} = \bm{I}$, and change the number of individuals and timesteps in $\{50, 100, 500, 1000\}$. In the second series of tests, the number of individuals and timesteps remains constant and equal to $100$ while the parameter that gives the importance of the contributions $\bm{\Sigma}_{\alpha}$ varies in $\{0, 0.5, 1.0, 1.5\}$. The case $\bm{\Sigma}_{\alpha} = 0$, in particular, is interesting, for it means that the model will see variables that do not contribute anything to the observed variables. In this case, if its performance is significantly lower than a model that does not have access to the variables, this would mean that the proposed model overfits the exogenous variables. In order to mitigate the randomness, we repeat the experiments ten times for each combination of parameters, generating a whole new set of parameters each time.

\paragraph*{Results} The results are summarized in table \ref{tab:results_synthetic}. As expected, the models which do not take into account the variables (\texttt{Clust+Seg}) or which try to fit a regression before any clustering (\texttt{Reg $\rightarrow$ Clust $\rightarrow$ Seg} and \texttt{Reg $\rightarrow$ (Clust+Seg)}) obtain ARI scores that are noticeably low. This is because the contributions we generated are independent of each other, meaning that fitting regression coefficients on the whole dataset always produces statistically insignificant parameters. 

The only case when these models are relevant is the case where $\bm{\Sigma}_{\alpha} = 0$, meaning that the exogenous contributions are always zero. For this experiment, the score only depends on whether clustering and segmentation are simultaneously carried out: \texttt{Reg $\rightarrow$ Clust $\rightarrow$ Seg} and \texttt{(Clust+Reg) $\rightarrow$ (Seg+Reg)} are both worse than the other three models, which perform similarly. As expected, the position of the linear regression in the estimation process has no significant importance for this case.

\begin{table*}[]
`\renewcommand{\arraystretch}{1.4}
    \centering
    \begin{tabular}{|c|c|c|c|c|}
    \hline
                                         &            \multicolumn{4}{|c|}{$\bm{\Sigma}_{\alpha} = \bm{I}$}     \\   \hline 
                                         &                     $I=T=50$         &                 $I=T=100$        &                 $I=T=500$        &       $I=T=1000$      \\  \hline
     \texttt{Reg $\rightarrow$ Clust $\rightarrow$ Seg}  & $ 0.334 \pm 0.152 $  &             $ 0.293 \pm 0.062 $  &             $ 0.286 \pm 0.061 $  &   $ 0.377 \pm 0.088 $    \\
     \texttt{(Clust+Reg) $\rightarrow$ (Seg+Reg)}  &       $ 0.589 \pm 0.141 $  &             $ 0.643 \pm 0.161 $  &             $ 0.620 \pm 0.081 $  &   $ 0.788 \pm 0.114 $  \\
     \texttt{Reg $\rightarrow$ (Clust+Seg)}          &       $ 0.463 \pm 0.099 $  &             $ 0.440 \pm 0.125 $  &             $ 0.414 \pm 0.112 $  &   $ 0.511 \pm 0.123 $   \\
     \texttt{Clust+Seg}                  &                 $ 0.478 \pm 0.106 $  &             $ 0.453 \pm 0.120 $  &             $ 0.431 \pm 0.120 $  &   $ 0.561 \pm 0.152 $  \\
     \texttt{Clust+Seg+Reg} (Proposed)   &       $ \mathbf{ 0.708 \pm 0.143 }$  &   $ \mathbf{ 0.784 \pm 0.110 }$  &   $  \mathbf{ 0.865 \pm 0.088}$  &   $  \mathbf{ 0.895 \pm 0.072 } $    \\ \hline \hline
                                         &            \multicolumn{4}{|c|}{$I=T=100$}     \\   \hline 
                                         &  $\bm{\Sigma}_{\alpha} = \bm{0}$ &  $\bm{\Sigma}_{\alpha} = (0.5^2)\bm{I}$ &  $\bm{\Sigma}_{\alpha} = \bm{I}$ &  $\bm{\Sigma}_{\alpha} = (1.5^2)\bm{I}$      \\   \hline 
     \texttt{Reg $\rightarrow$ Clust $\rightarrow$ Seg}      &   $ 0.568 \pm 0.178 $  &   $ 0.456 \pm 0.125 $  &   $ 0.346 \pm 0.063 $  &   $ 0.323 \pm 0.090 $    \\
     \texttt{(Clust+Reg) $\rightarrow$ (Seg+Reg)}  &   $ 0.542 \pm 0.177 $  &   $ 0.605 \pm 0.122 $  &   $ 0.661 \pm 0.174 $  &   $ 0.626 \pm 0.145 $  \\
     \texttt{Reg $\rightarrow$ (Clust+Seg)}          &   $ 0.645 \pm 0.165 $  &   $ 0.548 \pm 0.094 $  &   $ 0.467 \pm 0.121 $  &   $ 0.442 \pm 0.133 $   \\
     \texttt{Clust+Seg}                  &   $ 0.686 \pm 0.092 $  &   $ 0.586 \pm 0.137 $  &   $ 0.499 \pm 0.179 $  &   $ 0.434 \pm 0.107 $  \\
     \texttt{Clust+Seg+Reg} (Proposed)   &   $ \mathbf{ 0.689 \pm 0.104 }$  &   $  \mathbf{0.794 \pm 0.075 }$  &   $ \mathbf{ 0.815 \pm 0.088 }$ &   $  \mathbf{0.804 \pm 0.121} $    \\   \hline 
     \end{tabular}
    \caption{The mean and standard deviation of the Adjusted Rand Score obtained with ten generations of synthetic data. The method \texttt{(Clust+Seg) $\rightarrow$ Reg} is absent because it produces the same partition as \texttt{Clust+Seg}. The best result each time is in bold. }
    \label{tab:results_synthetic}
\end{table*}

\subsection{The Public Transport Ridership Dataset} \label{subsec:IdFM_dataset}

\begin{table}[]
    \centering
    \def\arraystretch{1.25}
    \begin{tabular}{|c|c|}
        \hline
        number of individuals  &   542 stations  \\ \hline
        days &   2,004  \\ \hline
        Measures per day &    1    \\ \hline
        \thead{covariates \\(number of variables)} &   \thead{type of transport available (4),\\ day of the week (7), lockdowns (3),\\strikes (2), holidays (2),\\year-periodic splines (20), time (1)}   \\ \hline
        \end{tabular}
    \caption{An overview of the real dataset.}
    \label{tab:dataset_summary}
\end{table}

To validate our model on real data, we used a recording of daily entries in the Parisian public transport system.\footnote{available at \url{https://data.iledefrance-mobilites.fr/explore/dataset/histo-validations-reseau-ferre/} and \url{https://data.iledefrance-mobilites.fr/explore/dataset/validations-reseau-ferre-nombre-validations-par-jour-1er-semestre/information/}} Gathered by the transport organization authority called Île-De-France Mobilités (IdFM), this dataset covers five and a half years, from January 1, 2017, to June 30, 2022. We also focus on the rail network (underground, public train, and tramway). Note that the records only cover the entries of travelers entering the network, which means that we have the origin of trips but not the destination. Note also that due to the organization of the ticketing system, one person may validate several times for a single trip when changing fare zones (if the journey includes connections between different modes). We do not try to compensate for this. Finally, we must emphasize that even though we use the term "individual" to designate each time series, one series of entries corresponds to a station, and not to a traveler.  
 
\paragraph{Preprocessing}
The raw data contain the number of entries for each type of ticket or subscription the travelers used. As we do not make use of this information, we simply consider the sum of all entries for each day and each station. However, only the couples (\textit{station, day}) with at least one entry are present. One might be tempted to say that an absence of data for a given day means that no user entered this station (which might have happened during the lockdowns, for instance), but we have no way of knowing which days are missing due to actual errors in the data collection process, and which days are absent because of the lack of users. Thus, we leave missing data as-is. 

To ensure anonymity, when a station recorded less than five entries for the same type of ticketing in a day, the dataset only contains the mention "less than five" for this type of ticket. In such cases, we considered that there were three entries for this day, station, and ticket type. We removed stations with more than 60\% missing days or less than 500 entries per day on average.
To handle the many noisy days in the data, we compute a two-week moving average of each station's series of entries. Each value that is below one-tenth of the result is removed.

Finally, given the strong imbalance between stations, we normalized the number of entries of each station. For each station, we divide the number of entries by the average number of entries during the first year, strikes excluded (from January 1, 2017 to December 31, 2017). We consider this period to be representative of usual behavior. 
Then, similarly to \cite{de_nailly_what_2021}, we took the log in base 10 of the number of entries.

\paragraph{Exogenous variables}
We include several dummy (binary) variables that relate to the type of day: one variable to know whether the day is a working day (meaning no holidays or weekends), seven variables for the days of the week, and two variables for strikes: as one strike was particularly between December 2019 and January 2020, we dedicate one variable to this period. The other 'strike' variable is the same for all other strikes. Similarly, we create three variables for the three lockdowns declared by the government. Finally, to account for tendencies throughout the year, such as a lower ridership during the summer, we use 20 year-periodic splines with degree 2. We also include four variables that are proper to each station, denoting the presence of four commercial types of train: underground (Paris only), RER (express regional network), transilien (a network going further than RERs), and commercial train (presence of national line in the same station).

\subsection{Experiments with real data} \label{subsec:results_real_data}

Section \ref{subsec:synthetic_data} demonstrated that the model is able to find the true cluster of individuals generated using Gaussian distributions. This section shows that even with real data, the model we propose outperforms the baselines defined in section \ref{subsec:synthetic_data}. 

However, for many real-life datasets, we have no way of knowing which individuals belong to each cluster, or even if there is a 'right' number of clusters. We first begin to select the number of clusters using the slope heuristic \cite{baudry_slope_2012}. As we expect the data to exhibit one change point at the beginning of the pandemic, we select $S=2$ segments \textit{a priori}. We begin by fitting the parameters for models with one to nineteen clusters. Each time, we obtain a (training) log-likelihood, which increases with the number of parameters. The slope heuristic consists in applying a penalty equal to twice the slope of the curve giving the log-likelihood as a function of the number of parameters in each model \cite{baudry_slope_2012}. In our case, with $S=2$ segments, the optimal value is reached for $K=5$ clusters (results not shown).

To evaluate our model on real data, we perform five-fold cross-validation, dividing the dataset into five subsets of mutually-independent samples. This constraint of mutual independence is not trivial: if one split the dataset into separate periods, for instance, the folds would not be independent. The reason is that the cluster assigned to each individual remains constant: if we know the segment a given day belongs to, we can infer the individual's cluster, which provides information about the cluster the individual belongs to outside of the training interval. This is why we perform cross-validation by dividing the dataset into five groups of individuals. We use the first four to estimate the parameters of the model, keeping the last group to measure the log-likelihood of the unseen data. We repeat the process four additional times, changing the validation dataset, and display the mean and standard deviation in table \ref{tab:results_real}. 

We can draw several conclusions from the results in table \ref{tab:results_real}: firstly, the two methods that perform clustering and segmentation separately are noticeably worse than the rest which means that performing clustering and segmentation at the same time is paramount. Secondly, the model that imposes the regression coefficients to be equal across clusters and segments (\texttt{Reg $\rightarrow$ (Clust+Seg)}) explains unseen data as well as a model that does not have access to the exogenous variables (\texttt{Clust+Seg}), hinting at overfitting. 

Finally, the best two models are the ones that both perform clustering and segmentation at the same time, while still allowing variable contributions to differ between segments (\texttt{(Clust+Seg) $\rightarrow$ Reg} and \texttt{Clust+Seg+Reg}). Between them, the proposed approach that performs clustering/segmentation and regression at the same time (\texttt{Clust+Seg+Reg}) is significantly better than the sequential model where the estimation of the contributions is done after the estimation of the partition ($p < 0.01$ using an unpaired t-test).  

However, this improvement comes at the cost of a higher computational burden: using a desktop computer\footnote{a computer with an Intel I7 @ 2.80GHz and a 15 Go RAM, along with Ubuntu 20.04}, the implementation that estimates all parameters at the same time with the EM algorithm takes fifteen minutes to converge on average ($15.4 \pm 2.25 $ min), which is four times longer to run than its sequential counterpart ($4.0 \pm 0.6 $ min for \texttt{(Clust+Seg) $\rightarrow$ Reg}).

Using cross-validation, we demonstrated that the proposed model is able to reach a higher log-likelihood than its counterparts, indicating a better ability to model the data. We will now use this model to understand the effects of the pandemic on public transport ridership. 



\begin{table*}[]
\renewcommand{\arraystretch}{1.4}
    \centering
    \begin{tabular}{c|c|c|c|c}
                      Model                              &                   validation LL                    & \thead{number of \\ parameters} & \thead{Segments and clusters\\ are computed ...} &  \thead{Regression \\ coefficients are...} \\  \hline
     \texttt{Reg $\rightarrow$ Clust $\rightarrow$ Seg}  &          $ -2.02 \times 10^6 \pm 3.0 \times 10^3 $  &  $73$ &  Sequentially    & Common\\ 
     \texttt{(Clust+Reg) $\rightarrow$ (Seg+Reg)}        &          $ -1.96 \times 10^6 \pm 8.2 \times 10^2 $  & $424$ &  Sequentially    & Different \\ 
     \texttt{Clust+Seg}                                  &          $  2.03 \times 10^6 \pm 5.6 \times 10^3 $  &  $34$ &  Simultaneously  & Absent \\ 
     \texttt{Reg $\rightarrow$ (Clust+Seg)}              &          $  2.05 \times 10^6 \pm 5.9 \times 10^3 $  &  $73$ &  Simultaneously  & Common \\ 
     \texttt{(Clust+Seg) $\rightarrow$ Reg}              &          $  2.86 \times 10^6 \pm 7.0 \times 10^3 $  & $424$ &  Simultaneously  & Different  \\ 
     \texttt{Clust+Seg+Reg} (Proposed)                   &   $ \mathbf{3.14 \times 10^6 \pm 8.2 \times 10^3 }$ & $424$ &  Simultaneously  & Different  \\ 
     \end{tabular}
    \caption{The mean and standard deviation of the log-likelihood computed on the validation dataset, using five-fold cross-validation. The best result is in bold. }
    \label{tab:results_real}
\end{table*}

\subsection{Analysis of the results on public transport ridership}

As mentioned at the beginning of the previous section, we apply the proposed model with five clusters and two segments. As the changes between segments are easier to explain than the difference between clusters, we will begin by explaining how the model provides insights into the public transport ridership before and after the pandemic, before using this behavior to understand how clusters differ from each other. 

\begin{figure*}
    \centering
    \includegraphics[width=15cm]{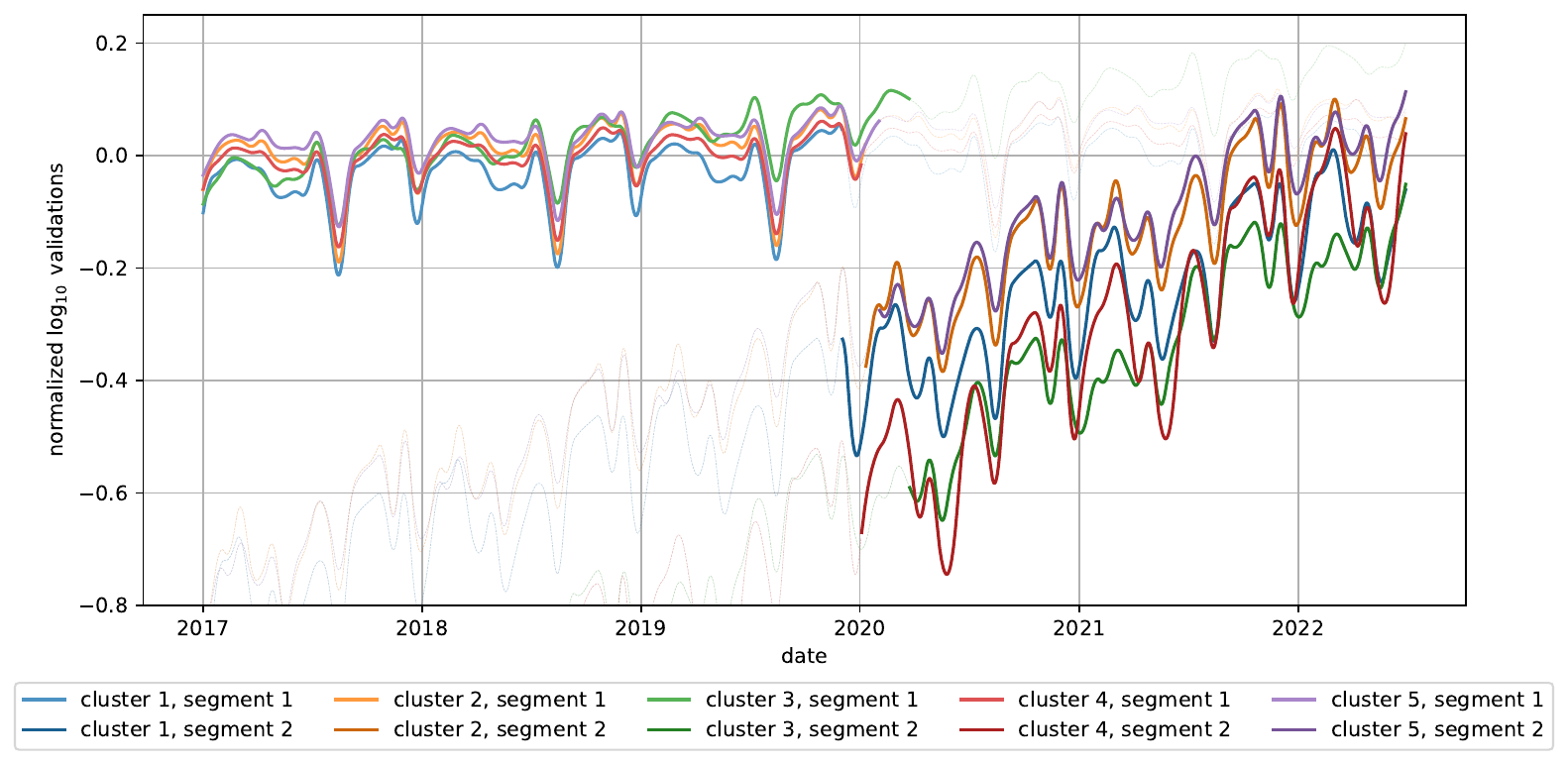}
    \caption{The splines associated with each cluster, along with the effect of time. Dashed lines denote the continuation of each segment in regions where it is unlikely.}
    \label{fig:idfm_reconstruction_splines}
\end{figure*}

Figure \ref{fig:idfm_reconstruction_splines} displays the sum of the mean and contributions of the time and the year-periodic splines. As expected, the mean of the second segment is noticeably lower than the first segment's, and the (positive) regression coefficient associated with the time models the recovery of each cluster. The impact of the pandemic on ridership is the gap between the reconstructions of the two segments, when their prior probability is equal (approximately early 2020, depending on the clusters). As expected, the gap between the extension of the first segment and the second segment is at its highest at the start of the pandemic, after which the number of entries increase to model the recovery of the network. 

Given the importance of the pandemic on the population's trips, one could expect the model to pick up the start of the pandemic (and the first lockdown) as the change point between segments. What happened is slightly different: a major strike occurred between December 2019 and January 2020. Even though we created a dummy variable to model the effect of this specific strike, one variable is not enough to model the continuous recovery of the signal. As a consequence, for some clusters, the model underestimates the effect of this strike, and assigns the lowest days to the next segment, the segment dedicated to the pandemic. This is why the transition between segments is sometimes located at the very beginning of the year 2020.

The small number of days assigned this way to the second segment is not, however, enough to prevent the model from accounting correctly for the events of the pandemic. In March 2020, the French government decreed a lockdown to fight against the pandemic. While people were allowed to travel to work, companies and institutions alike were encouraged to implement teleworking to avoid their employees taking public transport. After the end of the first lockdown, even though travel for personal reasons became allowed again, the teleworking measures remained in place to a certain extent and kept on affecting public transport ridership. This is what we want to measure with our model.

\begin{figure*}
    \centering
    \includegraphics[width=15cm]{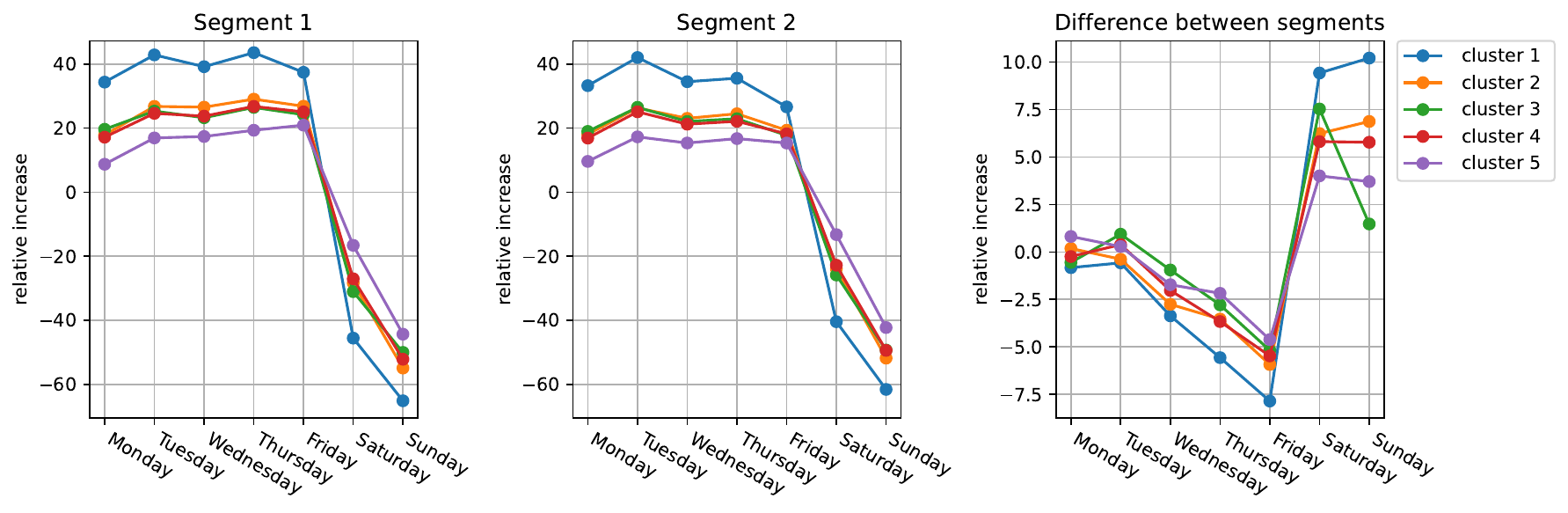}
    \caption{The effect of the seven days of the week on each of the two segments (left, center) and the difference in regression coefficients for each cluster (all values are translated from the log scale to a linear scale). We deal with the co-linearity by computing the coefficients such that the effect of the sum of the coefficients is absorbed by each segment's mean (which does not appear on this figure).}
    \label{fig:idfm_regression_coefficients}
\end{figure*}

Another source of information is the regression coefficients for each day of the week (fig. \ref{fig:idfm_regression_coefficients}). Both segment one (corresponding to normal behavior) and segment two (during and after the pandemic) have an increased number of entries during the weekdays and a decreased ridership on weekends.  
But when we look at the difference between the coefficients of two segments, we notice that in several cases, the coefficients associated with weekdays decreased with the pandemic and the weekend's coefficients increased. This means that the difference between weekdays and weekends is less noticeable after the pandemic. We hypothesize that this effect is due to the use of teleworking policies that reduced the number of commuters using public transport every day. Please note that fig. \ref{fig:idfm_regression_coefficients} is not affected by modal shift, as the effect of the sum of the seven covariates is absorbed by the mean ($\bm{m}_{k,s}$ in eq. \ref{eq:contribution_exogenous}) and not the variables encoding for the day of the week. 
The influence of the covariates varies from cluster to cluster, which will be helpful when understanding what makes clusters specific.

To know which features distinguish one cluster from another, the first method is to look at the regression coefficients (fig. \ref{fig:idfm_regression_coefficients}). We perform a t-test on the difference of regression coefficients across a couple of clusters to know which are the coefficients that change significantly between two clusters. To compare clusters, we focus on the first segment, corresponding to a more usual behavior. Although the entries of all clusters increase during weekdays and decrease during weekends, the amplitude of weekly variations differs between clusters which is informative to us.

The first cluster has the highest variations between weekdays and weekends, indicating a higher rate of commuters than the other clusters. It includes stations from the North, West, and South suburbs, as well as the center and West Paris boroughs (results not shown), two regions with moderate to high rates of employments-to-surface \cite{etienne_model-based_2014}. 
As we expected, the stations with the highest weekly variations are also those where this variation dampened the most after the pandemic: the stations with the most commute are also the ones where teleworking policies are the most impactful on public transport. 

The next three clusters have average weekly variations. Cluster two is, by all accounts, close to the average, except for its variance (see table \ref{tab:variances}): both its first and second segment have particularly low variances. This means that this cluster is dedicated to stations with little noise and average weekly variations. 

Conversely, cluster three has a noticeably high variance before the pandemic. This means that this cluster comprises all the stations where variations in the number of entries cannot be explained using the covariates. Surprisingly, variance in the second segment is moderate.  

Cluster four stands out for its high variance after the pandemic. It groups all stations for which the number of entries cannot be accurately explained using the covariates we provided (such as the official lockdowns, for instance). It should also be noted that the decrease in the number of entries in these stations is the largest out of all five clusters, along with cluster three (fig. \ref{fig:idfm_reconstruction_splines}). This is compensated by the highest recovery rate of all clusters: if we consider the impact of  time on the logarithm of the number of entries, the number of entries in the second segment increases by $74 \%$ per year on average. In comparison, the second segment of cluster three increases by $60 \%$ per year, while the other three clusters increase by $37 \%$ to $42 \%$ per year during the period following the pandemic. Thus, cluster four comprising all the stations whose series of entries were affected by the pandemic. 

Finally, cluster five has the smallest differences between weekdays and weekends of all the clusters. Like cluster two, it has a small variance, indicating that the stations in this cluster have variations which the model explains correctly using the exogenous variables. The decrease in the number of entries (fig. \ref{fig:idfm_reconstruction_splines}) is among the lowest of all clusters, along with cluster two. Geographically, the stations in this cluster are located almost exclusively in inner Paris (table \ref{tab:variances}), where housings are the most common \cite{etienne_model-based_2014}.

To sum up, the main element that distinguishes the clusters is the degree of variation between weekdays and weekends: clusters one and five have the largest and smallest variations, respectively. Among the middle three clusters, one of them (cluster two) had a small variance around an average behavior, while the remaining two had a high variance either before of after the pandemic. Surprisingly, no cluster had particularly high variance for both its segments. Additionally, one could have expected the consequences of the pandemic to be more diverse on the stations with high a commute, but the model groups all these stations into a single cluster.

\begin{table}[]
    \centering
    \begin{tabular}{c|c|c|c}
      Cluster   & \thead{Variance of\\segment one} & \thead{Variance of\\segment two} & \thead{Average distance to \\the center of Paris (\textit{km})} \\ \hline 
      1  & $1.8 \times 10^{-2} $   &  $5.4 \times 10^{-2} $    &  $16.5$ \\
      2  & $5.7 \times 10^{-3} $   &  $2.2 \times 10^{-2} $    &  $7.2 $\\
      3  & $3.5 \times 10^{-2} $   &  $5.1 \times 10^{-2} $    &  $16.0$\\
      4  & $7.9 \times 10^{-3} $   &  $1.7 \times 10^{-1} $    &  $11.8$\\
      5  & $4.8 \times 10^{-3} $   &  $1.7 \times 10^{-2} $    &  $ 5.8$\\

    \end{tabular}
    \caption{ The variance in each segment of each cluster, along with the average distance between the stations in the cluster and the Center of Paris. 
    }
    \label{tab:variances}
\end{table}

\section{Conclusion}\label{sec:conclusion}

Clustering, segmentation, and regression are three well-explored research problems, to the point the intersection between any two of them has been covered in the literature. We extended the clustering-segmentation model from \cite{same_model-based_2011}, and completed it by adding the contribution of exogenous variables. We developed the estimation of parameters using the Expectation-Maximization algorithm, and experimented with it on both synthetic and real data, demonstrating the interest in a joint estimation of parameters as opposed to a sequential one.


Using this model on a dataset of entries in the Paris public transport network allowed us to understand how the Covid-19 pandemic affected the public transport ridership. As expected, all stations saw the number of entries decrease sharply at the start of the first lockdown, with a slow recovery during the two years that followed. In addition,   
most clusters saw the number of entries further decrease during weekdays and increase during weekends, compared to their respective pre-pandemic levels. We attribute this effect to teleworking policies all the more so because the cluster which is defined by the highest proportion of commuters saw the strongest decrease in ridership during weekdays and the strongest increase during weekends. 
Whenever the variables we provided are not enough for the model, the model assigns a high variance to the concerned segment, as this was the case for another of the five clusters the model uncovered: the second segment of cluster four had a high variance, indicating that the evolution of the time series assigned to it is more irregular than the simple effects we included in with the covariates.

The next step would be to find a way to reduce the computational load of the proposed model. One could, for instance, think about initializing the EM algorithm with coefficients found from a simpler version of the model. Another possibility for improvement would be to use model selection criteria to choose the number of segments automatically, such as penalized criteria (BIC, AIC, \textit{etc.}) or dedicated heuristics \cite{baudry_slope_2012}.

\section*{Acknowledgments}

This work was supported by Vinci through the funding of the Eco Conception Chair in École des Ponts ParisTech.

\bibliography{references}
\bibliographystyle{abbrv}

\end{document}